\documentclass[%
 aip,
 amsmath,amssymb,
 reprint,%
]{revtex4-1}

\usepackage{amsmath}
\usepackage{siunitx}
\usepackage{mathtools}
\usepackage{graphicx}
\usepackage{dcolumn}
\usepackage{bm}

\usepackage[utf8]{inputenc}
\usepackage[T1]{fontenc}
\usepackage{mathptmx}
\usepackage{etoolbox}
\usepackage{float}
\usepackage{xcolor} 
\usepackage{graphicx}
\usepackage{subcaption}

\makeatletter
\def\@email#1#2{%
 \endgroup
 \patchcmd{\titleblock@produce}
  {\frontmatter@RRAPformat}
  {\frontmatter@RRAPformat{\produce@RRAP{*#1\href{mailto:#2}{#2}}}\frontmatter@RRAPformat}
  {}{}
}%
\makeatother
\begin{document}

\preprint{AIP/123-QED}
\title{\centering A compact setup for $^{87}\text{Rb}$ optical tweezer arrays}
\author{Xue Zhao}
\affiliation{%
School of Physics and Key Laboratory of Quantum State Construction and Manipulation (Ministry of Education), Renmin University of China, Beijing 100872, China}
\author{Xiao Wang}
\affiliation{%
School of Physics and Key Laboratory of Quantum State Construction and Manipulation (Ministry of Education), Renmin University of China, Beijing 100872, China}
\author{Wentao Yang}
\affiliation{%
School of Physics and Key Laboratory of Quantum State Construction and Manipulation (Ministry of Education), Renmin University of China, Beijing 100872, China}
\author{Xiaoyu Dai}
\affiliation{%
School of Physics and Key Laboratory of Quantum State Construction and Manipulation (Ministry of Education), Renmin University of China, Beijing 100872, China}
\author{Yirong Wang}
\affiliation{%
School of Physics and Key Laboratory of Quantum State Construction and Manipulation (Ministry of Education), Renmin University of China, Beijing 100872, China}
\author{Guangren Sun}
\affiliation{%
School of Physics and Key Laboratory of Quantum State Construction and Manipulation (Ministry of Education), Renmin University of China, Beijing 100872, China}
\author{Fangshi Jia}
\affiliation{%
School of Physics and Key Laboratory of Quantum State Construction and Manipulation (Ministry of Education), Renmin University of China, Beijing 100872, China}
\author{Kuiyi Gao}
\thanks{Author to whom correspondence should be addressed: kgao@ruc.edu.cn}
\affiliation{%
School of Physics and Key Laboratory of Quantum State Construction and Manipulation (Ministry of Education), Renmin University of China, Beijing 100872, China}
\author{Wei Zhang}
\thanks{wzhangl@ruc.edu.cn}
\affiliation{%
School of Physics and Key Laboratory of Quantum State Construction and Manipulation (Ministry of Education), Renmin University of China, Beijing 100872, China}
\affiliation{%
Beijing Academy of Quantum Information Sciences, Beijing 100093, China}

\date{\today}

\begin{abstract}
We describe a simple and compact experimental setup for optical tweezer arrays of $^{87}$Rb atoms. This setup includes a compact vacuum system, a single cooling laser, a simple tweezer laser, and a flexible control system. The small vacuum system with only 40 cm length takes advantage of the high atomic flux two-dimensional magneto-optical trap (2D MOT) while maintaining a low background pressure in the 3D MOT chamber ensuring sufficient lifetime of the trapped atoms. Atom number of the laser cooled sample of $\sim 2 \times 10^7$ and temperature of $ 92\ \mu\text{K}$ is achieved. The flexible control system with real-time waveform generator modules (RWG) provides precise control of all the RF devices, and enables real-time feedback control of both the global and individual beams in optical tweezer arrays. An optical tweezer array with $25 \times 25$ homogeneous traps is demonstrated. This simple and compact demo setup makes it more accessible to experimental quantum physics. 
\end{abstract}

\maketitle

\section{Introduction}

Optical traps using the AC Stark shift to exert strong light forces on small particles are widely applied to trap atoms for quantum simulation\cite{Greiner2002,Bloch2008,Browaeys2020}, quantum computing\cite{Henriet2020,Bluvstein2024} and quantum metrology\cite{Young2020,Kaufman2021}. Typically, strongly focused dipole traps are used to trap bulk gases of atoms, while optical lattices from beam interference and optical tweezer arrays from light diffraction create a large-scale ordered light patterns to trap atomic many body systems for quantum experiments\cite{Bloch2008,Browaeys2020}. In the past decade, optical tweezer arrays, as a bottom-up approach to prepare a scalable and programmable quantum systems of neutral atoms, have brought experimental quantum science into a new era\cite{Browaeys2020,Young2020,Morgado2021}.

To realize an optical tweezer array of cold atoms or molecules, one usually needs a complicated setup\cite{Endres2016,Barredo2016,Chiu2025,Manetsch2025,Norcia2019,Jenkins2022,Zhang2025a,Schaeffner2024,Eckner2025,Bloch2023,Gruen2024} with a sophisticated vacuum system, multiple lasers of different wavelengths and precise multi-channel control system. An ultra-high vacuum system provides an experimental environment for atoms by isolating the atoms from atmosphere. A near-resonant cooling laser system with multiple lasers is used to collect and cool the atoms down to micro-Kelvin temperatures and imagine the atoms. A few far-detuned tweezer lasers and high-resolution objectives are also needed to trap, move, and manipulate the atoms. Additionally, a control system makes precise adjustments to all external fields for the atoms, including partial pressure of atoms in vacuum, laser beam intensities and frequencies, magnetic field values and gradients, real-time positions and powers of the optical tweezers, on a micro-second time scale. In the past, most experiments incorporated more complex setups for different applications\cite{Sheng2022,Wei2026,Anand2024,Anderegg2019,Holland2023,Pichard2024,Zhang2025b,Jin2026,Desiree2026,Schlosser2023}, while it is still worth some effort to simplify the optical tweezer experiment while keeping its features for many non-trivial applications.

Here, we describe a particularly simple and compact setup for optical tweezer arrays of $^{87}\text{Rb}$ atoms. This setup includes a compact vacuum system, a single cooling laser system, a simple tweezer laser, and a flexible control system. The vacuum system with only \SI{40}{\centi\meter} in length takes advantage of the rather high atomic flux of the 2D MOT to load the 3D MOT while keeping a low background gas pressure for a sufficient lifetime of the trapped atoms. The single \SI{780} {\nano\meter} cooling laser provides about \SI{2}{\watt} of power in total for cooling, repumping, probing, and frequency stabilizing, with all the beams delivered to the experiment by fiber-coupled devices. A laser cooled sample of atom number of $\sim 2\times10^7$ and temperature of $92\ \mu$K is obtained. The simple \SI{852} {\nano\meter} tweezer laser provides up to \SI{10}{\watt} power to the acousto-optical deflector (AOD) for optical tweezer arrays. Optical tweezer arrays with $25\times25$ homogeneous traps is demonstrated. The flexible control system not only outputs the conditional analog and digital control signals but also incorporates real-time waveform generator module (RWG) to make precise control of all the RF devices. Of particular importance, this RWG module will make it possible for real-time feedback control of both the global and individual beams in optical tweezer arrays. This setup with a compact MOT and flexible tweezer arrays makes various experiments in quantum physics more accessible.

\section{Experimental setup}
\begin{figure*}[htbp]
    \centering
    \includegraphics[width=18cm]{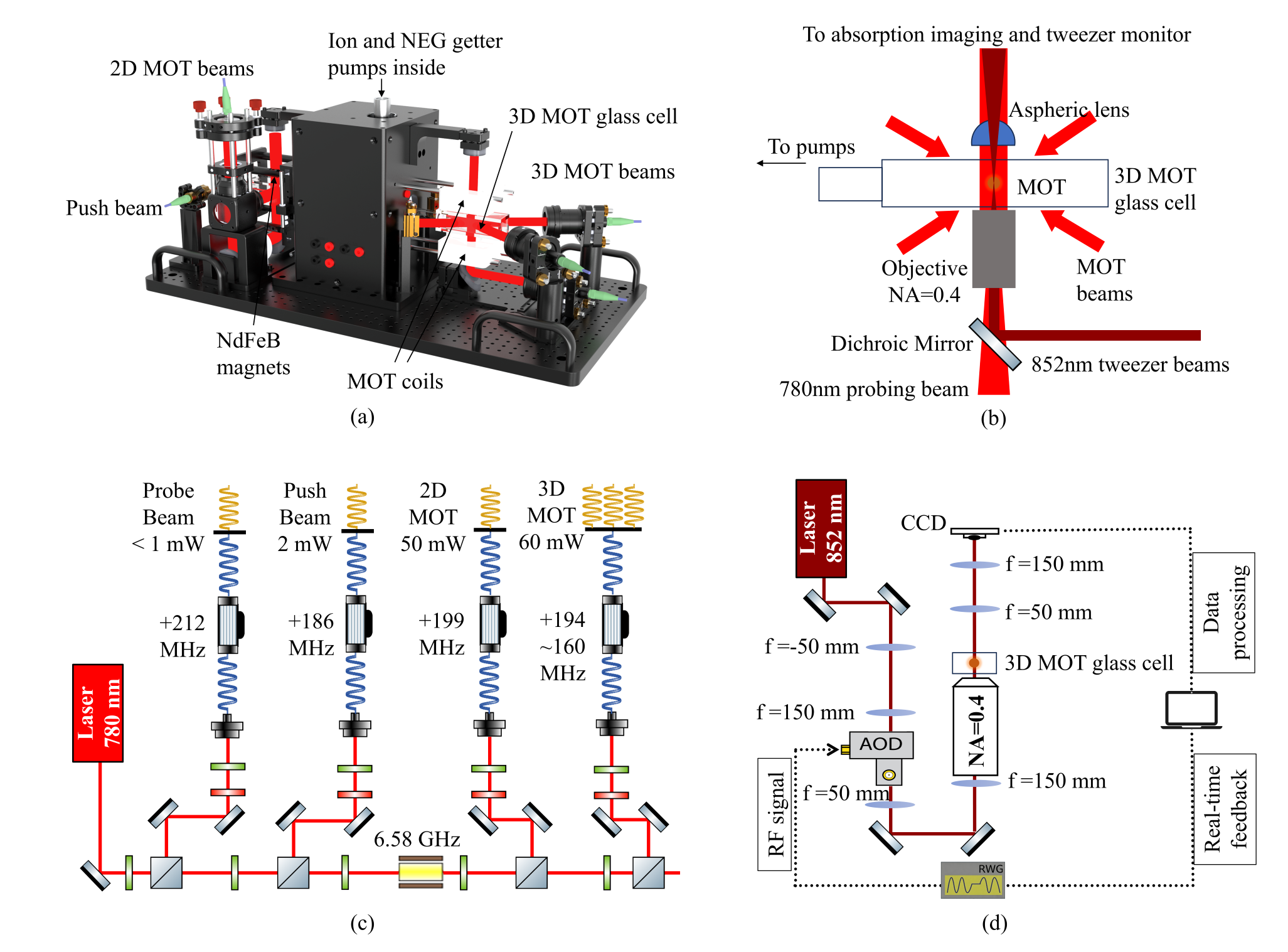}
    \caption{The $^{87}\text{Rb}$ optical tweezer arrays setup. (a) The vacuum system and laser beams of the MOT. (b) 3D MOT glass cell with imaging and optical tweezer beams. (c) Cooling laser system at 780 nm. (d) Optical tweezer laser system at 852nm.}
    \label{fig:four_images}
\end{figure*}

    

    

\subsection{Setup Overview}
Our setup to realize optical tweezer arrays of $^{87}$Rb atoms addresses three considerations: (1) a simple vacuum system of a MOT with both fast loading and long trapping lifetime simultaneously; (2) a simple laser system to provide sufficient power at different frequencies for cooling, repumping, and absorption and fluorescence imaging; (3) a relatively simple optical tweezer system with only one laser, one AOD and an objective.


\subsection{Vacuum system}
To achieve a fast loading of the 3D MOT for the subsequent trapping of the optical tweezer beams, a 2D MOT is chosen to provide sufficiently high atomic flux to the 3D MOT without degrading its background vacuum. As shown in Fig. 1(a), the vacuum system mainly consists of two glass cells, one uncoated 29$\times$29$\times$74 mm$^{3}$ glass cell for the 2D MOT and another anti-reflection coated glass cell of the same size for the 3D MOT. The total length of the vacuum system is \SI{40}{\centi\meter} along the axis of the glass cells. The 3D MOT cell is connected to a small ion pump and a non-evaporable getter (NEG) pump, while the 2D MOT cell is pumped by two NEG pumps.

A differential pump tube with a diameter of 2 mm between the glass cells maintains an ultra-high vacuum for the 3D MOT section while the partial pressure of $^{87}\text{Rb}$ atoms in the 2D cell varies by orders of magnitude for sufficient atomic flux. Therefore, it naturally satisfies the requirements of both fast loading and long lifetime of the atoms at different stages of the experiment. The atomic flux is determined by the current of the dispenser, which is set to 4.3 A during experimental operation. It can be easily adjusted for higher atomic flux and shorter MOT loading time. The dispenser will be replaced by isotopically pure $^{87}\text{Rb}$ source to increase the flux by approximately four times in the future.

\begin{figure*}[htbp]
    \centering
    \includegraphics[width=17cm]{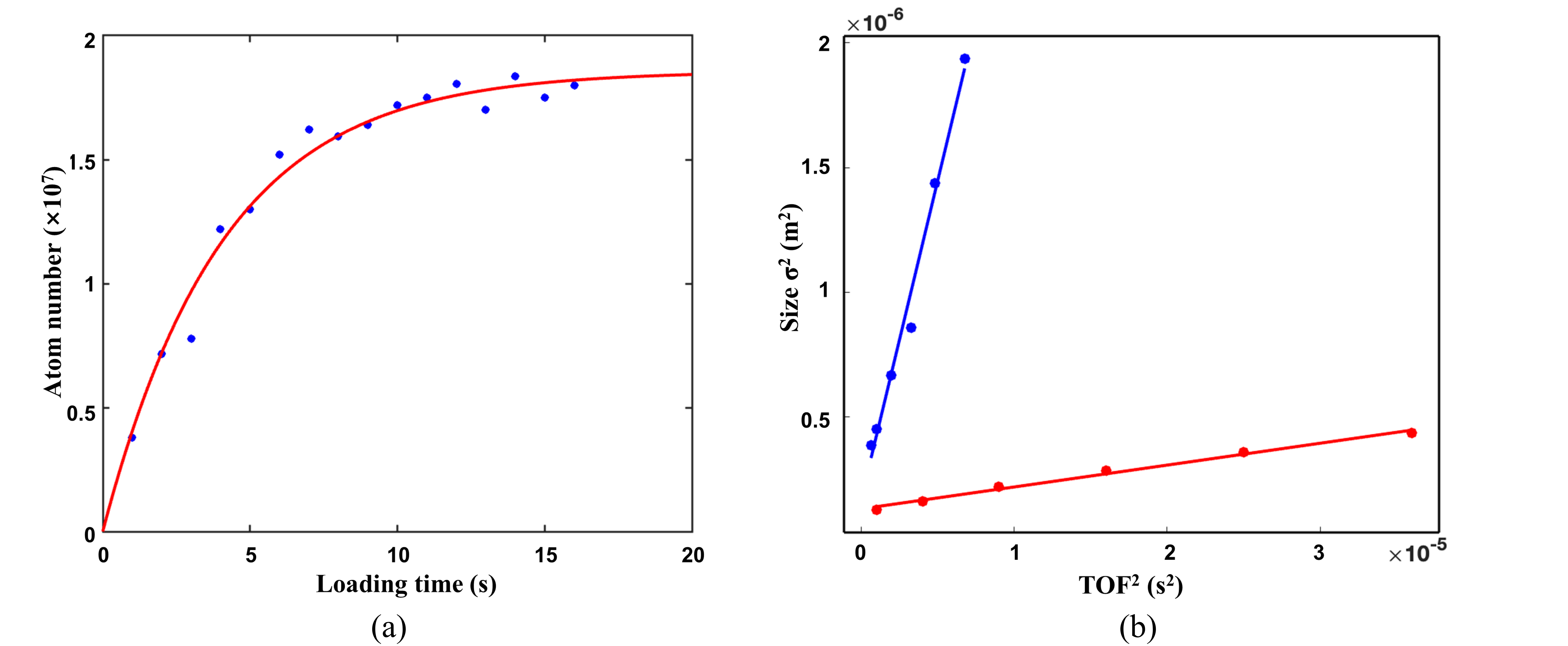}
    \caption{The $^{87}\text{Rb}$ MOT and molasses. (a) The atom number with varying loading time of the 3D MOT. The exponential fit gives a time constant of 4 s and a loading rate of $4.6 \times 10^6 \ \text{s}^{-1}$. (b) The size of the MOT and molasses with different time of flight. The linear fit indicates the temperature of the MOT is $\ 2.4\ \text{mK}$, while the temperature of the molasses drops to $\ 92\ \mu$K. }
    \label{fig:two_images}
\end{figure*}
\subsection{Laser system}
The laser systems involve a near-resonant 780 nm cooling laser for collecting, cooling, and probing the atoms, and a far-detuned 852 nm tweezer laser for trapping and manipulating the cold atoms.

As shown in Fig. 1(c), the 780 nm laser (Precilaser, FL-SF-780-2-CW) is frequency-stabilized to the crossover signal of D2 transitions by modulation transfer spectroscopy. It provides a maximum output of 2 W and the beam is split by a waveplate and PBS into different paths for spectroscopy, cooling and probing the atoms. The beams for the 2D MOT, 3D MOT, pushing and probing are sent to fiber-AOMs then delivered to the vacuum system. The repumping component of the 2D and 3D MOT is generated by the 6.58 GHz electro-optic modulator (EOM) before the fiber AOMs.

The 852 nm laser (Precilaser, FL-SF-852-10-CW) is a free-running laser for trapping and manipulating the atoms in the optical tweezer arrays, since its frequency is far detuned from the transitions of $^{87}\text{Rb}$ atoms. As shown in Fig. 1 (d), it provides a total power of up to 10 W, and its output beam is expanded to a diameter of 2 mm, then sent to a 2D AOD to generate hundreds of beams before being focused into micrometer size by an aspherical lens.
As shown in Fig. 1 (b), the 780 nm probing beam is combined with the 852 nm tweezer beam by a dichroic mirror. A telescope equiped with a high-resolution objective (NA=0.4) demagnifies and projects hundreds of tweezer beams with waists down to $\sim 1.2\ \mu\text{m}$ to trap the atoms in the glass cell while simultaneously collimating the probing beam to obtain absorption images of the atoms. Meanwhile, the same objective also collects the fluorescence from the atoms to form fluorescence images, especially for atoms individually trapped in the tweezer arrays. The distances and powers of the tweezer beams are precisely controlled by the RF signals sent from our control system to the AOD.

            
            
    

\begin{figure*}[htbp]
    \centering
    \includegraphics[width=13cm]{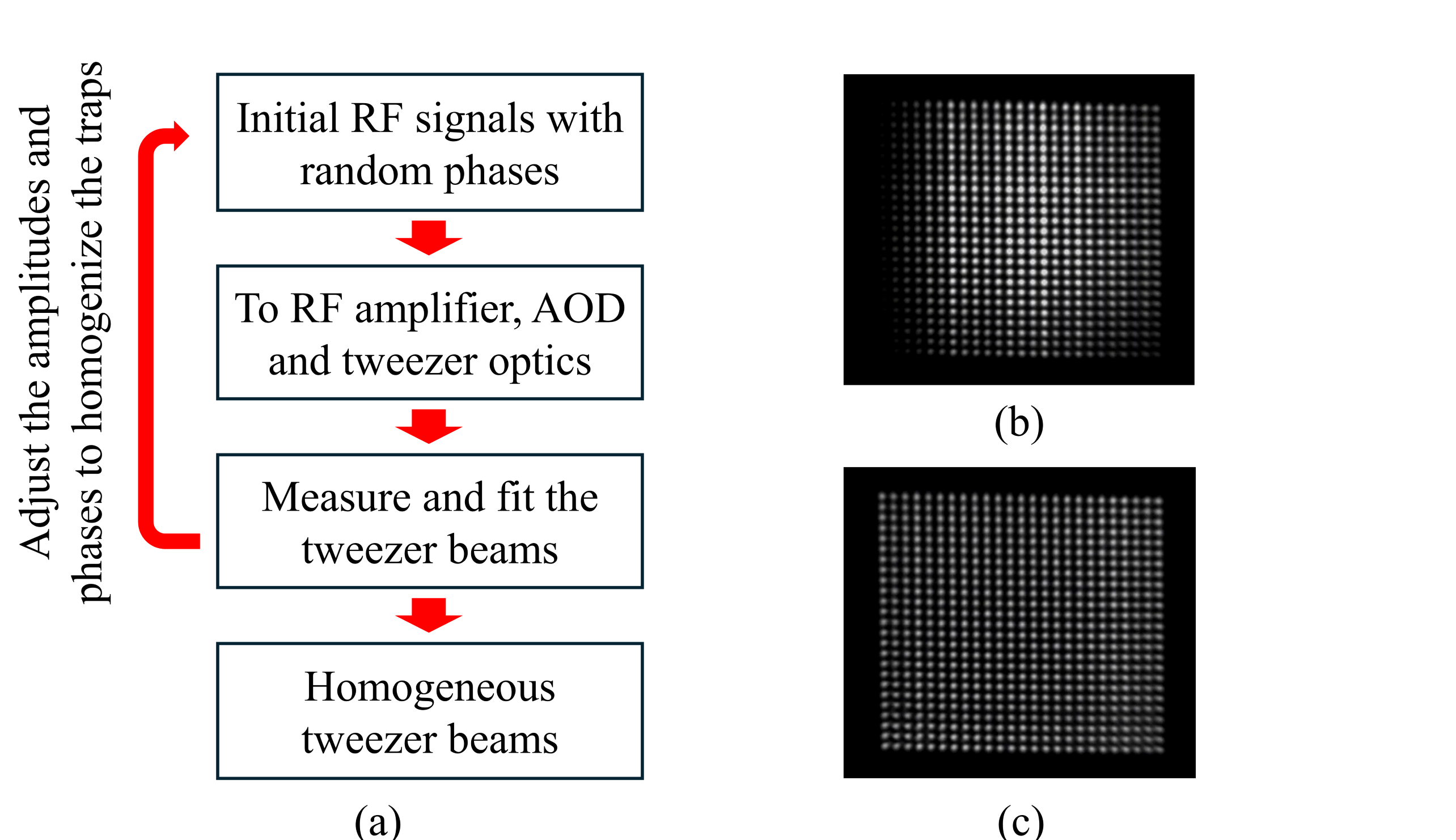}
    \caption{\raggedright Optical tweezer arrays.(a) Optimization of the tweezer beams.(b) imagine of the 2D tweezer arrays without optimization.(c) a 2D optical tweezer arrays with $25 \times 25$ homogeneous traps.}
    \label{fig:three_images}
\end{figure*}

\subsection{MOT}
As shown in Fig.~\ref{fig:four_images}, the 2D MOT beam is frequency shifted by a fiber-AOM with 199 MHz and delivered to 2D MOT optics with a total power of $\sim 60$ mW. The collimated beam is then split into two by a PBS and sent in the horizontal and vertical directions, respectively. The retro-reflected beams provide transverse cooling of the atoms in the 2D MOT glass cell. The 2D quadrupole magnetic field is created by four strong grade N52 neodymium magnets each with dimensions of 5 mm$\times$3 mm$\times$60 mm. The magnets are mounted on the vacuum chamber with a separation of 70 mm between them, and the differential pump tube is positioned at the center of the magnets. The push beam is frequency shifted by 186 MHz by a fiber-AOM and precisely aligned to the axis of the differential pump tube.

The MOT beam is shifted by a fiber AOM with 194 MHz and sent to a 1-to-3 fiber splitter, where the total power of the beams is more than 50 mW. The three output beams are collimated to form three pairs of retro-reflected MOT beams, each with a diameter of 15 mm. They are used to collect and cool the Rb atoms in the 3D MOT glass cell. The horizontal beams intersect at $120^\circ$ to avoid being clipped by the high-resolution objective and imaging optics. The quadrupole magnetic field for the MOT is generated by a pair of coils in an anti-Helmholtz configuration. The coils have a diameter of 30-52 mm and consist of 100 turns wound from 1 mm enameled wire. The coils create a magnetic field gradient of $B'_z$=27 G/cm at 3 A. The coils are controlled by an analog programmable power supply (Delta Elektronika, ES015-10). There are also three pairs of rectangular compensation coils around the glass cell to cancel for the ambient magnetic field.

\subsection{Control system}
The control system (Future Instrument, WYS8U) provides flexible control signals, including not only traditional analog and digital output signals for precise timing and control of different instruments in the lab, but also an RWG module with frequency up to 400 MHz, to control the RF devices such as AOM and AOD. This RWG module enables real-time control of all individual beams in the optical tweezer arrays. In particularly, this system provides fast feedback control of RF signals with a time delay $\sim500 ns$, which may be crucial to many challenging manipulations of the trapped atoms.

\section{Laser cooling}
The experiment usually starts with the 2D and 3D MOT. When the dispenser, the MOT beams and the magnetic gradient field are turned on, the 2D MOT provides the transversely cooled atomic flux to load the 3D MOT through the differential pump tube. The atom number of the 3D MOT is obtained by the absorption imagining of the atomic cloud. As shown in Fig.~\ref{fig:two_images}(a), the atom number of the 3D MOT increases exponentially as the loading time increases. Under the optimal magnetic field gradient and detuning of the laser beams, the time constant is 4 s and the saturated atom number is $\sim 2 \times 10^7$ with an initial loading rate of $4.6 \times 10^6 \ \text{s}^{-1}$. The temperature of the atomic cloud is obtained by analyzing the Gaussian size of the cloud as a function of time of flight. Under optimal conditions, the temperature of the $^{87}$Rb atoms in 3D MOT is $\ 2.4\ \text{mK}$, as shown in Fig.~\ref{fig:two_images}(b).

In order to load the laser-cooled atoms into the optical tweezer arrays, an optical molasses is used to further cool the atoms after the MOT. For $^{87}$Rb atoms, polarization gradient cooling (PGC) is applied when the magnetic gradient field is turned off right after the MOT loading process. The compensation fields in all directions must be precisely adjusted by changing the currents through the compensation coils. During the 16 ms PGC, the detuning of the cooling beams is ramped from -15 MHz to -48 MHz and the intensity is also ramped down to 6 mW. After the PGC, the temperature of the atoms drops significantly to $\ 92\ \mu\text{K}$, as shown in Fig.~\ref{fig:two_images}(b).

\section{Optical tweezer arrays}
The optical tweezer arrays are created by projecting and focusing the multiple diffracted beams from the AOD. Thus, the positions and distances of the beams are precisely controlled by the RF signals. In our case, the RF signals used to drive the AOD and control the optical tweezer beams are synthesized by the RWG module of our experimental control system. For multiple optical tweezer beams, it is natural to send RF signals with many different frequency components, so that the diffraction angles and positions of the tweezers can be precisely controlled. Using this approach, static or moving tweezer arrays are obtained. However, the nonlinear response of the imperfect RF amplifier and AOD leads to unexpected mixing signals at the sum and difference of the initial input signals from the RWG. One can randomize the phases of different frequency components, calculate the mixing output of different frequency components, and then optimize the phases of the initial RF signals, to minimize the sum of all difference tones to suppress the unexpected interference signals\cite{Endres2016}, which would otherwise lead to an inhomogeneous distribution of the tweezer arrays.

A different strategy is chosen here in our experiment. We treat the optical tweezer system, including the RF amplifier, the AOD and the tweezer optics, as a black box. In general, the amplitudes and phases of the RF signals input from the RWG can be optimized by analyzing the intensity distribution of the tweezer arrays. The initial amplitudes of the multiple frequency components with the same distances are roughly adjusted by compensating for the difference in diffraction efficiency of the AOD. The phases of the signals are initially randomized. This leads to an inhomogeneous intensity distribution of the tweezer arrays, which is measured by a monitor camera. Then one can change the amplitudes and the phases of the signals one at a time to reduce all the unexpected tweezer beams caused by the frequency mixing and finally achieve a homogeneous intensity distribution across all intended tweezers. As shown in Fig.~\ref{fig:three_images} (a), the whole optimization process of the tweezer arrays can be implemented in four steps: 1. Initialize the amplitudes and randomize the phases of the RF components. 2. Measure the light distribution and fit each tweezer beam with a Gaussian function to get the intensity distribution of the traps. 3. Adjust the amplitudes of the RF frequency components one by one to minimize the mixing tweezer beams. 4. Iterate the above steps until homogeneous optical tweezer arrays are obtained. Fig.~\ref{fig:three_images} (b) shows the 2D tweezer arrays without optimization, while a 2D optical tweezer arrays with 25$\times$25 homogeneous traps after optimization is demonstrated in Fig.~\ref{fig:three_images} (c).

\section{Conclusion and outlook}
We report a simple and compact setup for optical tweezer arrays of $^{87}\text{Rb}$ atoms. This setup includes a compact vacuum system, a single cooling laser, a simple tweezer laser, and a flexible control system. A laser-cooled sample with an atom number of approximately $2 \times 10^7$ and a temperature of approximately $92\ \mu\text{K}$ is obtained. An optical tweezer array with $25 \times 25$ homogeneous traps is demonstrated. The capability provided by the real-time waveform generator module can enable an unprecedented real-time feedback control of both the global tweezers and the individual traps, which paves the way for fast control of the trapped atoms in future experiments. This simple and compact demo setup makes experimental quantum physics more accessible.

\begin{acknowledgments}
We acknowledge financial support by the National Natural Science Foundation of China (Grants No. 12274460, 12074428, and 92265208), and the National Key R\&D Program of China (Grant No. 2022YFA1405301). Furthermore, we thank Future Instrument for providing the control system for our experiment.
\end{acknowledgments}

\section*{Author declarations}
\subsection*{Conflict of Interest}

The authors have no conflicts to disclose.

\section*{Data Availability}
The data that support the findings of this study are available from the corresponding author upon reasonable request.

\bibliography{aipsamp}

\end{document}